\documentclass{optica-article}

\journal{opticajournal}

\articletype{Research Article}

\begin{document}

\title{Design and Fabrication of Ultrahigh Q Chip-Based Silica WGM Micro-resonators for Single-Atom Cavity-QED}

\author{Tal Shahar Ohana\authormark{1 + *}, Gabriel Guendelman\authormark{1 +}, Eran Mishuk\authormark{2 +}, Nadav Kandel\authormark{1}, Dror Garti\authormark{1}, Doron Gurevich\authormark{1}, Ora Bitton\authormark{2} and Barak Dayan\authormark{1}}

\address{\authormark{1}Department of Chemical Physics, Weizmann Institute of Science, Rehovot, Israel\\
\authormark{2}Department of Chemical Research Support, Faculty of Chemistry, Weizmann Institute of Sciences, Rehovot, Israel\\
\authormark{+}Equal contribution} 

\email{\authormark{*}tal.ohana@weizmann.ac.il}

\begin{abstract*} 
Of the many applications of whispering-gallery mode (WGM) microresonators, Single-atom cavity-QED poses the most extreme demands on mode-volume and quality factor. Here we present a model-based procedure for the fabrication of small mode-volume ultrahigh Q silica WGM microresonators of varying geometries, from toroidal to micro-spheres. We experimentally demonstrate WGM resonators with ultra-high qualities as high as $1.7\times 10^8$ at 780nm. We present a theoretical model that allows tailoring the recipe to attain the desired geometry of the fabricated WGM microresonators.
\end{abstract*}

\section{Introduction}
Optical whispering-gallery mode (WGM) micro-resonators \cite{braginsky1989quality, fused-silicaResonators1, vernooy1998high} are a vital tool for a vast variety of applications,
from bio-sensing and particle detection \cite{sensing2,sensing3,rosenblum2015cavity}, through nonlinear optics \cite{Torroidal_Spillane} and lasers \cite{DemonstrationVahala, comb, FrequencyComb1, FrequencyComb2}, to cavity quantum electrodynamics (cQED) \cite{haroche2014exploring,lefevre1997towards,aoki2006observation}. In particular, high-quality and small mode-volume are important for maximizing the Purcell enhancement \cite{purcell1995spontaneous} required for coupling such resonators to single quantum emitters \cite{electrodynamics,cavity_QED_Kimble,aoki2006observation}. Among all these applications, single-atom cQED poses the strongest demands for high intrinsic quality factor (Q), combined with the smallest possible mode volume. This is crucial for several reasons. First, to facilitate single-photon quantum operations, such as photon-atom gates \cite{reiserer2014quantum, bechler2018passive},linear losses must be minimized and cannot be compensated for by gain or increased intensity. Second, unlike sensing applications, the WGM microresonator is maintained in a vacuum, so its Q factor is not constrained by the surrounding medium. Instead, it is limited by intrinsic imperfections, such as material loss, surface roughness, and bending-induced radiation loss, which effectively establish a lower limit for the radius and, consequently, the mode volume of these microresonators.

Within the large variety of micro-resonators, silica (SiO$_2$) based WGM micro-resonators are well established platforms  and have received considerable attention  \cite{Vahala2003}. Optical WGMs are dielectric round structures in which light is confined by continuous total internal reflection along the periphery of the micro-resonator in a whispering gallery mode, which is named after its acoustic equivalent. 
Coupling of light to WGMs is performed evanescently by using adjacent thin waveguides or tapered optical nanofibers. The monolithic geometry of WGM resonators makes them highly resilient to vibrations.
By attaining high surface quality and the low intrinsic material losses, WGMs reach ultra-high quality factors of $Q>10^8$ \cite{armani2003ultra,lee2012chemically,Chen2017}. Additionally, WGMs are localized close to the rim of the resonator, which allows interactions with quantum emitters or  other media within the WGM's evanescent field. Both the WGM resonance and its coupling to the waveguide can be readily controlled thermally and by varying the distance between the WGM and the waveguide, respectively. 

Lastly, WGM microresonators fabricated on silicon chips are more compatible with scalable silicon-photonics fabrication processes than Fabry-Perot resonators, and allow direct integration with photonic chips\cite{Chang:19}. 

In 1989, Braginsky and co-workers were the first to realize that nearly perfect spheres can be formed by melting high grade silica glass fibers \cite{braginsky1989quality}. The formation of spheres through surface tension (that
is, as a molten droplet) provided a near atomically smooth surface
with minimum scattering from the boundary roughness, a property that allowed record-high Q factors. 
In 2003, Vahala and co-workers managed to combine the benefits of surface tension induced smoothness and lithographic production by creating microtoroids out of silica on silicon with a very high-Q of $\sim10^8$ \cite{armani2003ultra}. This high-Q value represented an improvement of nearly four
orders of magnitude over previous chip-based resonators. Later, they demonstrated Kerr-nonlinearity induced optical parametric oscillation in such a high-Q micro-toroid \cite{Torroidal_Spillane}, as well as the achievement of extremely small mode volumes, reaching a $Q/V$ ratio of $\sim2.5\cdot10^6$ with the highest value of quality factor $Q\sim4\cdot10^8$ observed for microtoroids at telecom wavelengths\cite{kippenberg2004demonstration}. 
The high value of $Q/V$ is what makes toroidal resonators an extremely promising platform for cQED in the regime of strong coupling \cite{electrodynamics}.
The disk geometry presents several advantages over previous toroidal geometries as it enables direct use of ion implantation, a planar technology, for doping optical microcavities. 
Vahala and co-workers demonstrated an erbium-doped microdisk on-chip resonator with $Q\sim5\cdot10^7$ \cite{DemonstrationVahala}. Such chip-based silica toroidal microresonators have also been used for optical frequency comb generation \cite{comb} and the demonstration of octave-wide comb operation with microresonator with $Q\sim2.7\cdot10^8$ at $1\mu m - 2.15\mu m$ \cite{FrequencyComb1,FrequencyComb2}. 

Although the majority of works with dielectric WGM microresonators have been performed at telecom wavelengths, silica WGM microresonators support a broad range of wavelengths, including the visible and near infra-red wavelengths that are required for interaction with atomic systems. Indeed, Vahala and co-workers demonstrated high $Q$ for toroidal cavities at a wavelength of 852 nm, suitable for cQED with atomic cesium \cite{PhysRevA.71.013817}. Zhu et al. demonstrated detection and sizing of individual poly-sterene nanoparticles using scattering-induced mode splitting of a WGM in an ultrahigh-Q microtoroid at $\lambda=670 nm$ \cite{sensing2}.
Lastly, silica WGM micro-sphere and micro-toroid resonators have enabled the demonstration of on-chip coupling between single photons and single atoms, including photon-atom two-qubit gates \cite{shomroni2014all}, \cite{rosenblum2016extraction}, \cite{bechler2018passive}.

Here we report on the realization of high-quality toroidal and spherical resonators-on-a-chip with record-high Q factors of up to $Q=1.7\cdot10^8$ at 780nm. We describe in detail the fabrication procedure and present a numerical model that allows tailoring of various fabrication parameters, such as the etching steps, the CO$_2$ laser intensity and thickness of SiO$_2$ layer, to attain pre-determined microresonator geometries. 

\section{Fabrication Process}
Fabrication of silica WGMs micro-resonators on a silicon chip is a two-step process involving standard microelectronic technologies (i.e. photolithography) and reflow treatment of silica with CO\(_2\) laser \cite{jager2011high, righini2011whispering}.
It involves selection and tuning of several parameters: The silica’s thickness, disk diameter, pillar size and laser intensity. The fairly large number of parameters enables one to fabricate specific geometries, in several different ways. Importantly, choosing a certain set of initial conditions will define the final form of the resonator.
In this section, we provide a detailed, step-by-step guide to the fabrication process of the WGM microresonators.

\subsection{Photolithography and etching steps}
\begin{figure}[h!]
\includegraphics[width=0.80\textwidth]{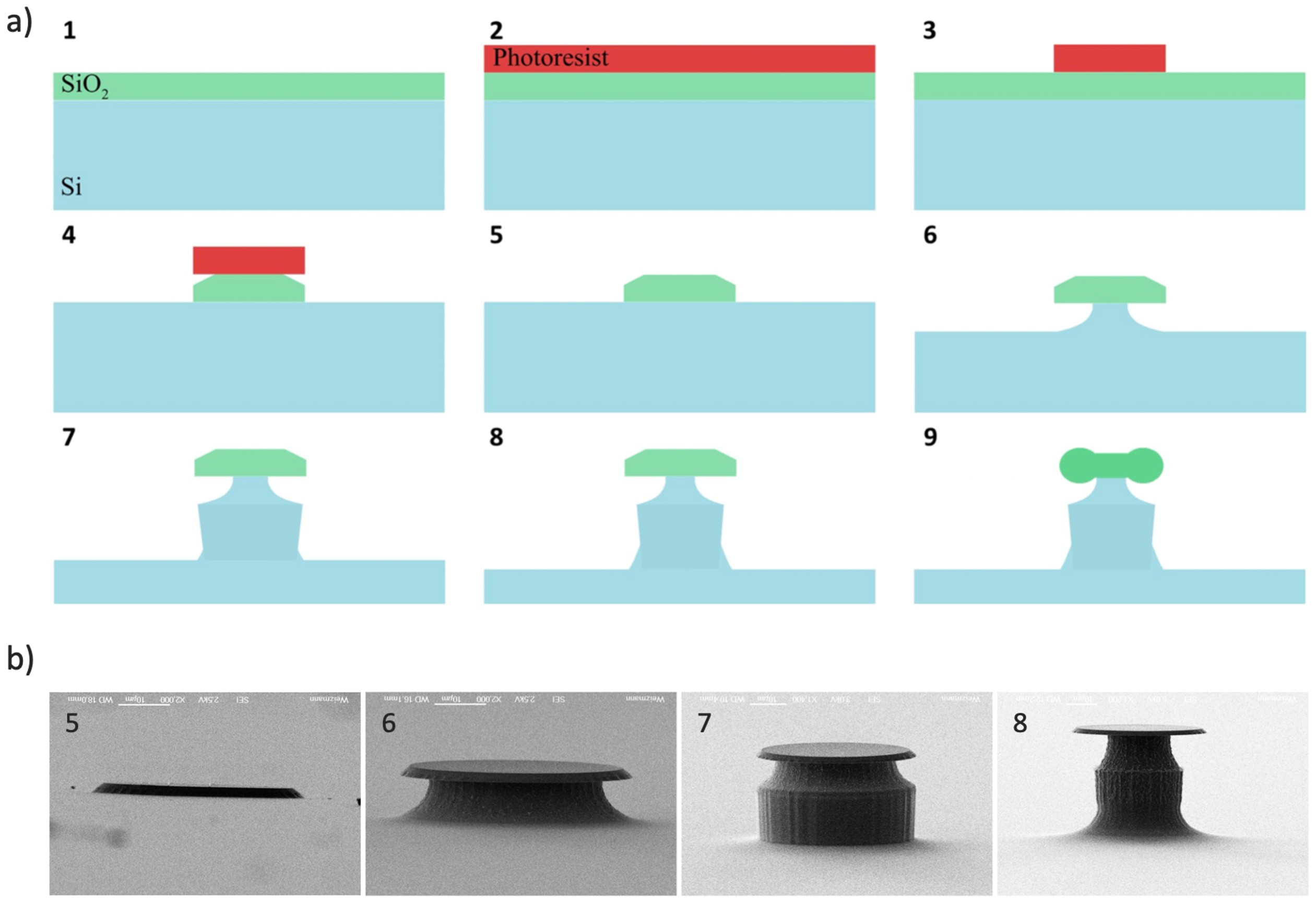}
\centering
\caption{Illustration of the complete nanofabrication process of WGMs micro-resonators. \textbf{a)} Photolithography and etching steps. \textbf{b)} SEM images of the resulting structures of the dry etching steps 5-8.}
\label{nanofab}
\end{figure}

The fabrication process steps, corresponding to the numbers shown in Figure  \ref{nanofab}, are outlined as follows:
As a first step, circular photoresist pads with a predetermined radius are patterned by optical lithography on a silica layer (steps 1-3), which was grown by wet or dry thermal oxidation of \(<\)100\(>\) silicon substrate. Next, the silica layer is wet-etched by a buffered oxide solution (step 4). Due to the isotropic nature of this process a wedge shape occurs. The wedge angle can be modified by varying the photoresist-silica adhesion or, in a more controllable way, by performing additional slow wet etching \cite{lee2012chemically}. After the wet etching step the remaining photoresist is thoroughly removed by using series of organic solvents followed by O\(_2\) and Ar plasma for the removal of renaming organic contaminates. Further cleaning with Piranha solution can be done as well if needed (step 5). The Si is etched beneath the SiO\(_2\) disk using a dry etching procedure (figure \ref{nanofab}b). It is done isotropically by using a standard ICP machine with a SF\(_6\) plasma (steps 6,8). An additional dry etching is done anisotropically to create a taller pillar (step 7).This anisotropic etching is optional and can be employed to increase the pillar height, to ease nanofiber coupling, while maintaining a large and more stable pillar.  Following is the reflow of the silica disks by an intense CO\(_2\) laser (step 9).

\subsection{CO$_2$-laser reflow and  evaporation of the silica}
\begin{figure}[h]
\centering
\includegraphics[width=\textwidth]{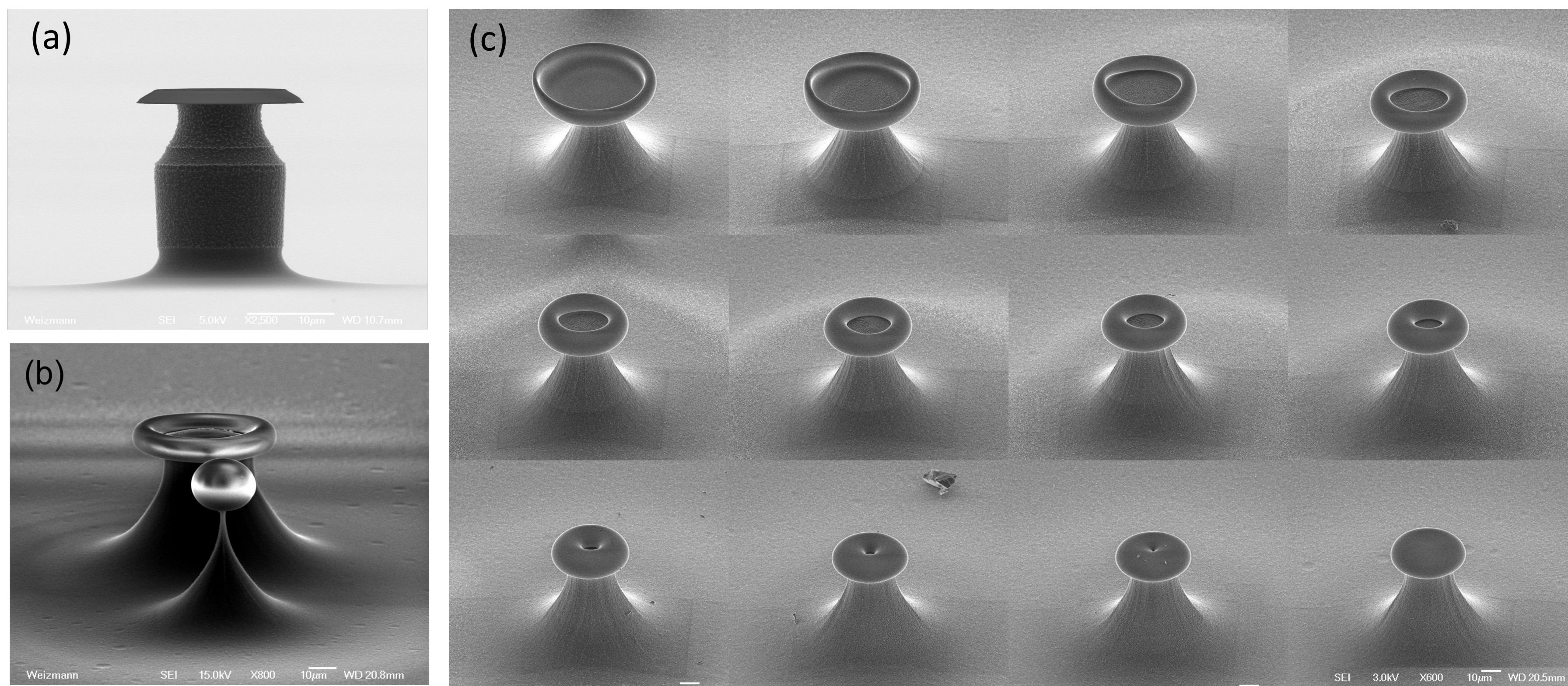} 
\caption{(a) SEM of the silica disk suspended on a silicon pillar as the initial geometry. (b) SEM image of sphere and toroid micro-resonators, with different pillar size, reflowed with a fixed CO$_2$ laser intensity. (c) SEM image view of micro-resonators with fixed geometrical parameters and different CO$_2$ laser intensity of the reflow process. Decreasing volume results from silica evaporation.}
\label{fig:tal}
\end{figure}

The reflow of the silica disk is done with a CO$_2$-laser (SYNRAD 48-1kwL). The beam is aligned such that the disk is fairly centered in its waist as shown in figure \ref{fig:tal}a. For the sake of homogeneous illumination and repeatability, the waist of the beam needs to be significantly larger than the disk (in our case, for \(100 \mu m\) disks, we used a Gaussian beam with divergence of \(\sim300 \mu m\)).
Since silica has high absorption at the CO$_2$-laser lines (10,200 - 10,600 nm), the irradiation heats it to temperatures where the viscosity is sufficiently low and it can flow spontaneously due to surface tension ($\sim$ 2400$^{\circ}$C) \cite{boyd2012surface,elhadj2012evaporation,kitamura2007optical,markillie2002effect,yang2010comparing}. The silicon pillar serves as a heat sink that eventually cools down the reflowed silica, and leads to the formation of a solid micro-resonator. As we show in figure \ref{fig:tal}(b) and detailed in the next sections, depending on the laser intensity, different initial dimensions of the silica disk and the silicon pillar will reflow to form different shapes which minimize the final surface energy. At high laser intensities, the silica's high temperature can result in non-negligible evaporation and consequent volume reduction (up to tens of percent). This ultimately affects the eventual shape of the resonator, as governed by the minimization of surface energy.
After the new microresonator geometry is formed, continued heating with the CO$_2$-laser leads to additional evaporation of silica from the structure, resulting in further re-shaping of the microresonator.
In figure \ref{fig:tal}(c) below we demonstrate the shape evolution of the micro-resonators by varying the reflow intensity (increasing from right to left) for constant geometrical parameters. As evident, increasing the laser intensity results in the melting and reflow of larger parts of the silica disk, together with evaporation of some of the silica, reducing the resonator overall volume.\\
Lastly, we performed a detailed analysis of surface morphology by atomic force microscope (AFM), which showed RMS roughness of less then 5 \AA, in agreement with previous works \cite{vernooy1998high}. We did not observe any meaningful correlation between the resonator's geometry and the surface roughness.

\begin{figure}[h!]
    \centering
    \includegraphics[width=1\textwidth]{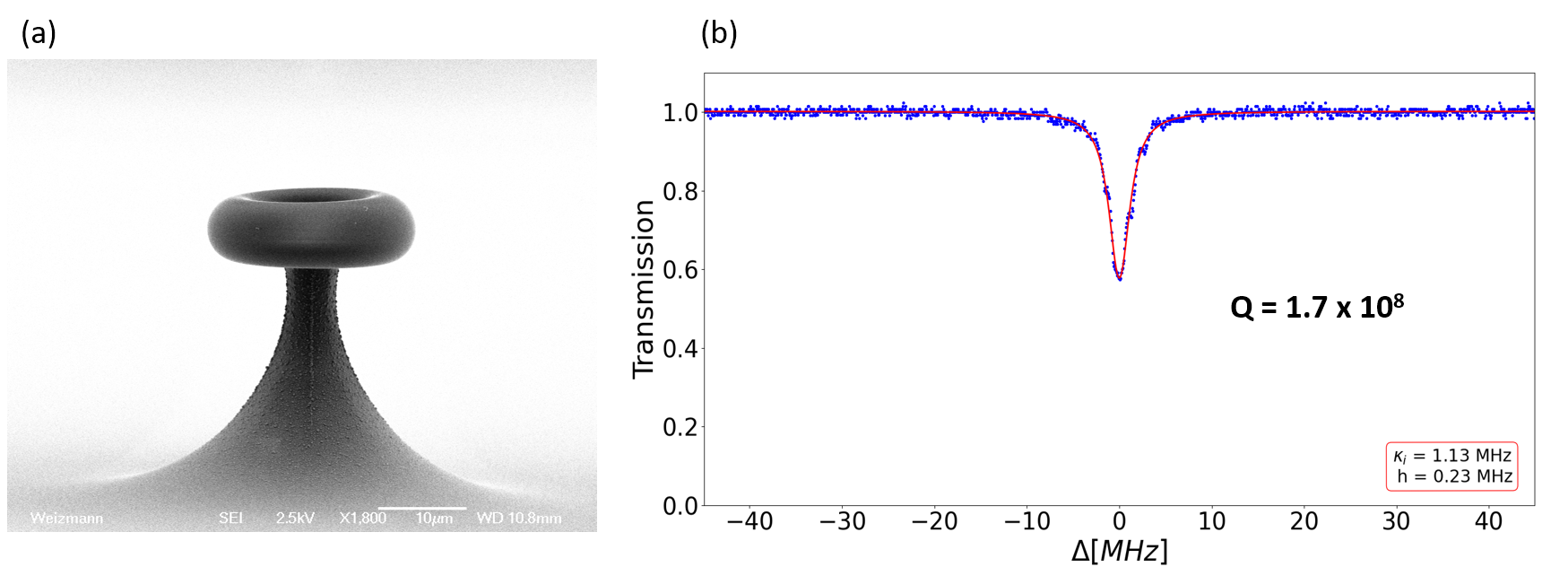}
    \caption{a) SEM image of an example toroidal WGM micro-resonator and b) its spectrum, corresponding to Q of \(1.7\cdot10^8\). The resonators in the UHV chamber have Qs that vary from \(3\cdot10^7\) to \(6\cdot10^7\)}
    \label{fig:resonator and spectrum}
\end{figure}

Following the above-mentioned fabrication procedure, we present in figure \ref{fig:resonator and spectrum}  one of our best fabrication result of an on-chip WGM micro-resonator, featuring quality factor of \(1.7 \cdot 10^8\).

\section{Predicting the final geometry of a WGM resonator}
Constructing a fully detailed model of the reflow dynamics is a formidable task. Instead, we outline a simplified model that nonetheless captures the main aspects of the process, and can be used to understand the physical parameters that govern the formation of the WGM microresonators, and to relate the initial fabrication parameters to the final resonators shape.
In the reflow process, the final geometry depends on the initial dimensions of the silica disk and the silicon pillar, and on the intensity of the CO$_2$ laser pulse. Mostly, it is a self-limiting process that reaches a near steady state once all the silica that was heated above the mobility temperature has reflowed into a structure with a lower cross-section and cooled down back below the mobility temperature. 

Specifically, the self-termination results from the exponential dependence of the viscosity on the temperature.
The flow of silica occurs at its mobility temperature ($T_{mobility}\sim$2400$^{\circ}$C, to reach viscosity of 10$^3$ Pa $\cdot$ s \cite{boyd2012surface,elhadj2012evaporation,kitamura2007optical,markillie2002effect,yang2010comparing}). To reach this temperature the CO$_2$ laser's heating rate must be greater than the heat dissipation rate, $\dot{Q}_{in} > \dot{Q}_{out}$, where heat dissipates mainly via conduction through the pillar.
At high laser powers, the silica heats up quickly to its mobility temperature and begins to fold inward from the edges toward the center. This folding reduces the cross-sectional area exposed to the CO$_2$ laser beam, which in turn decreases the heating rate. However, heat dissipation, which depends on the temperature difference between the heated silica and the cooler silicon pillar as well as the heat conduction to the pillar, continues to increase. 
At the point where $\dot{Q}_{in} < \dot{Q}_{out}$ the silica quickly cools down below its mobility temperature, practically stopping the reflow.
Any further shaping of the microresonator by the CO$_2$ beam is then the result of the slower and more gradual processes such as evaporation and creeping of the silica.
The model separates the shaping of the WGM resonator into two main parts. The first is the prediction of the mobile portion of the silica disk, and second is the folding to the final shape, driven by surface energy minimization.

\subsection{Determining the Radius of Mobility}

The combined effect of the illumination of the entire silica disk by the CO$_2$ laser versus its cooling via the silicon pillar results in a radial heat distribution, being highest at the edges and lowest at the center.
Assuming radial symmetry, we connect between the laser intensity and the outermost radius of the final resonator shape $r_{final}$ by defining the radius of mobility ($r_{mob}$) as the radius which separates the inner immobile region, where $T(r)<T_{mobility}$, from the outer mobile region, where $T(r)>T_{mobility}$. 
Given the sharp dependence of the mobility on the temperature, we approximate that beyond this radius, the silica is sufficiently mobile to fold,, whereas below $r_{mob}$ it remains completely immobile. 
For low laser intensities, only the edges of the silica disk, which are the furthest from the pillar, will heat enough to become mobile and start folding due to surface tension. The final outer radius of the resulting structure, $r_{final}$, will then be only slightly smaller that of the initial disk, $r_{disk}$. At high laser intensities nearly the entire disk will become mobile, except for the area above the pillar, leading to significantly smaller $r_{final}$. At extremely high intensities, and for long illumination time (several milliseconds), considerable evaporation of the silica can occur, resulting in $r_{final}$ which is only slightly larger than $r_{pillar}$ (see Fig. \ref{fig:tal}). \\
The radius of mobility is found by solving the time-dependent heat equation in radial coordinates, assuming axial symmetry and accounting for all the possible heat dissipation channels of conduction, radiation, convection and evaporation.
Since the thickness of the silica disk is small (2$\mu$m in our case) compared to the other dimensions, we neglect the heat distribution along the height axis. In addition, provided the silica disk is fairly aligned with the center of the much larger laser beam, we assume a uniform illumination in our model.

Figure \ref{fig:reflow_scheme} below illustrates the geometry of our system and the heating and dissipation rates associated with the laser reflow process.
\begin{figure}[h!]
\includegraphics[width=0.4\textwidth]{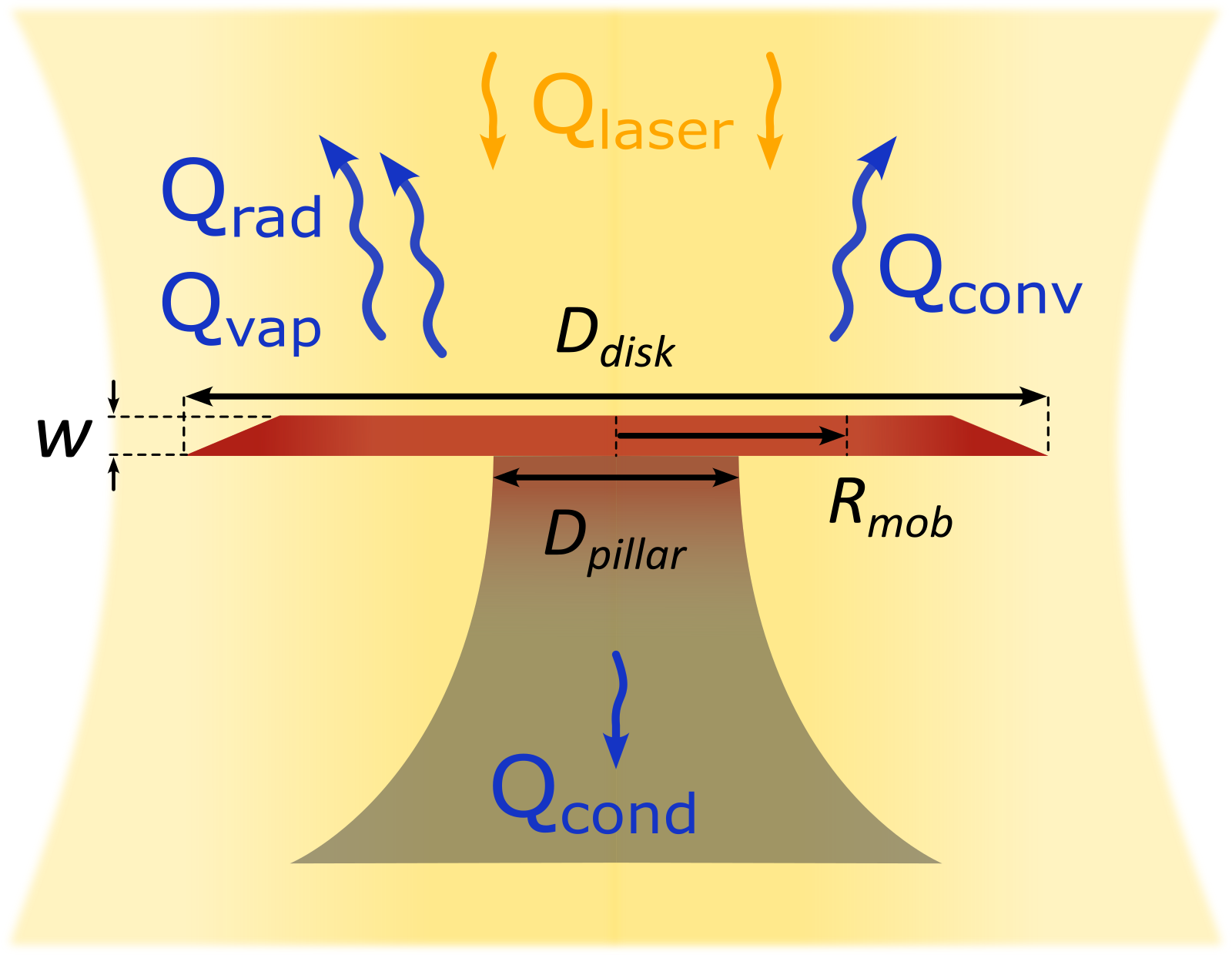}
\centering
\caption{Sketch of the reflow process of SiO$_2$ disk on Si pillar, specifying geometrical parameters and heat flows in the system.}
\label{fig:reflow_scheme}
\end{figure}
We denote the internal heat energy per unit volume of the silica disk at each point and time as \(Q=Q(r,t)\). The rate at which the internal heat energy per unit volume changes, \(\frac{dQ}{dt}\), is proportional to the rate at which the temperature changes, \(\frac{dT}{dt}\):
\begin{equation}
    \frac{ dQ}{dt} = C_p \rho \frac{dT}{dt},
\end{equation}
where \(C_p\) is the specific heat capacity of silica and \(\rho\) is the silica's density. \\
The energy balance in our system is:
\begin{equation}
    \frac{dQ}{dt} = -\frac{d}{dr}(-\dot{q}_{laser} + \dot{q}_{conduction} + \dot{q}_{radiation} + \dot{q}_{convection} + \dot{q}_{evaporation} )
\end{equation}
    \begin{align}
\dot{q}_{laser} &= w \alpha \frac{P_{laser}}{\mu ^2 _{beam} }\\
    \dot{q}_{conduction} &= w \frac{1}{r} \frac{\delta}{\delta r} (r\cdot k \frac{\delta T}{\delta r})\\
    \dot{q}_{radiation} &= \epsilon \sigma( T^4 (r) - T^4_{ambient})\\
      \dot{q}_{convection} &= h_{conv} (T(r)-T_{ambient})\\
         \dot{q}_{evaporation} &= H_{vap} f(T(r))
\end{align}

\noindent where \(f(T(r))\) is an Arrhenius expression fit for measured evaporation rates of bulk fused silica with activation energy of 120.1 kcal/mol taken from \cite{elhadj2012heating}:
\begin{align}
    &f(T(r)) = 6.25\cdot10^{-6} \exp{\left(-\frac{120.1}{RT(r)}\right)}     &\left[\frac{Kg}{\mu m^2 sec}\right]
\end{align}
\(w\) is the thickness of the silica disk, \(\alpha\) is the absorption coefficient of the silica, \(\frac{P_{laser}}{\mu^2_{beam}}\) is the heat flux of the laser beam, \(k\) is the silica's conduction coefficient, \(\epsilon\) is the emissivity of the silica surface, \(\sigma\) is Stefan-Boltzmann constant, \(T_{ambient}\) is the temperature of the surrounding, \(h_{conv}\) is the convection coefficient of the silica and \(H_{vap}\) is the enthalpy of the silica. Average values are taken for the silica's coefficients. \\
Numerically solving the heat equation, we find how the temperature is radially distributed, and thus the steady-state value of \(r_{mob}\) (typically reached within less than 5 ms) for each laser power, \(r_{pillar}\) and \(r_{disk}\).\\ 
Additionally, in order to calculate the evaporated volume, we roughly choose time of evaporation of \(\sim\)3ms. After this time we assume the structure has already cooled down to a temperature that corresponds to negligible evaporation rate. The evaporated volume is calculated as follows,
\begin{equation}
  V_{vap} = \int_{r_{pillar}}^{r_{disk}} \int_{0}^{2ms} 2\pi r \frac{dw(r)}{dt} dr dt
\end{equation}
where,
\begin{equation}
    \frac{dw(r)}{dt} = \frac{f(T(r))}{\rho} = 6.25\cdot10^{-6} \exp{\left(-\frac{120.1}{RT(r)}\right)}\frac{1}{\rho}
\end{equation}
Having found \(r_{mob}\) and \(V_{vap}\) from the dynamics of the heat equation, we can calculate the volume that becomes mobile during the reflow process, and from mass conservation considerations we can evaluate the final outer radius $r_{final}$ of each resonator geometry. For more details regarding the mass conservation of each geometry see the supplemental information.\\
In the following section we consider the minimization of the surface energy, and together with the results from the heat equation, we establish a general phase diagram of the possible geometries of the WGM microresonators.

\subsection{Surface energy minimization}
The driving force of the reflow process is surface tension, which aims to lower the surface energy. Essentially this means that all the silica that was heated enough to become mobile will reflow to form the shape that has the minimal surface energy. As this reflow typically creates a structure that is more compact than the original disk, with a significantly smaller cross-section with the CO$_2$ laser, the reflow stops when the silica cools down and "freezes" in the new shape. Assuming some $r_{mob}$ in the range between $r_{pillar}$ and $r_{disk}$, we can find which shape gives the minimal surface energy. For this we first need to calculate the surface areas, curvatures and volumes of the possible final shapes.
The initial geometry is a silica disk of radius $r_{disk}$ and width $w$, suspended on a silicon pillar of radius $r_{pillar}$. Thus, the initial volume of the silica is: 
\begin{equation}
V_{disk} = \pi r_{disk}^{2}w
\end{equation}
Assuming the silica volume right on top of the pillar remains cold and immobile:
\begin{equation}
V_{immobile} = \pi r_{pillar}^{2}w
\end{equation}
\begin{figure}[h]
\centering
\includegraphics[width=0.7\linewidth]{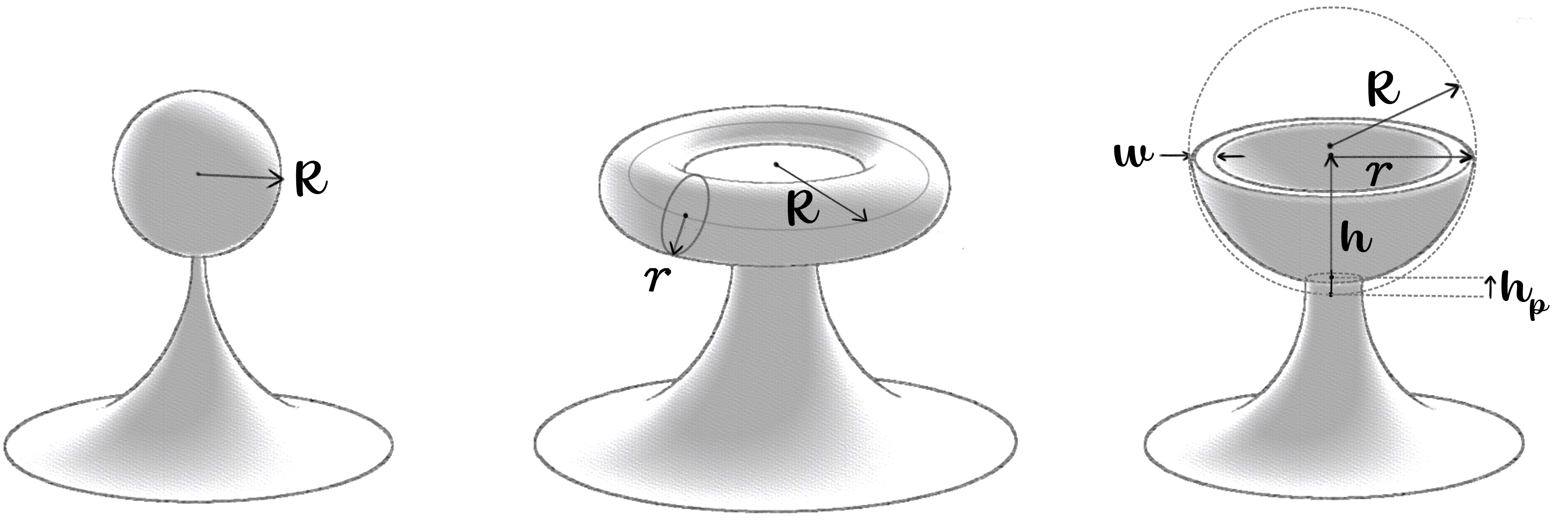}
\caption{\label{fig:final shapes} Final possible resonator shapes. From left to right : Spheroid, Torus, Cup}
\end{figure}
The final three general shapes that can generally be formed in the reflow process are a "cup", a torus and a spheroid (as shown in figure \ref{fig:final shapes}). We begin by calculating the volume of the cup shape:
\begin{equation}
V_{cup}(R) = \frac{\pi w}{3}\left( 6h(R)R - 3\left( R + h(R) \right)w + 2w^{2} \right) + \pi h_{p}^{2}(R)w\
\end{equation}
where $h(R)$ is the height of the cup and $h_p$ is the height at which the pillar is supporting the cup. both heights are measured from the most bottom point on a hypothetical sphere of radius $R$ from which the cup was derived (see figure \ref{fig:final shapes}). $R$ can be expressed in terms of the cup's parameters, 
\begin{equation}
R = \frac{r^{2} + h^{2}}{2h}
\end{equation}
the radius of curvature of the cup is: 
\begin{equation}
    \frac{1}{R_{cup}} = \frac{1}{R}\left(1+\frac{w}{2R}\right) 
\end{equation}
\begin{equation}
V_{torus} = \left( \pi r^{2} \right)(2\pi R) + \pi(R - r)^{2}w\
\end{equation}

where $r$ and $R$ are the minor and major radius of the torus.
The torus radius of curvature is: 

\begin{equation}
    \frac{1}{R_{torus}} = \frac{1}{2}\left(\frac{1}{R} + \frac{1}{r}\right)
\end{equation}

For the sphere shape, we allow a certain amount of ellipticity, as observed in our experiments. In this case, the volume is:

\begin{equation}
V_{spheroid} =\frac{4}{3}\pi R^{2}r = \frac{4}{3}\pi R^{2}\sqrt{R^{2}-r_{pillar}^{2}}
\end{equation}

where the amount of ellipticity is defined by flattening parameter, 
\begin{equation}
f=\frac{R-r}{R} 
\end{equation}

and the radius of curvature is:
\begin{equation}
\frac{1}{R_{spheroid}}=\frac{1}{R} + \frac{1}{r}
\end{equation}

Using the volume expressions maintained for the cup, torus and spheroid, we numerically solve the mass conservation equation:
\begin{equation} \label{eq:mass conservation}
    V_{initial} - V_{final} = (V_{disk} - V_{immobile}) -  V_{cup/torus/spheroid} = 0
\end{equation}
 by solving equation \eqref{eq:mass conservation}, for each shape we get sets of radius of curvature and surface area values. 
 The surface areas are expressed as follows:
 \begin{equation}
     A_{cup} = 2\pi (2R(h-h_p)-w(R+h-h_p-w)) + \pi r_{pillar}^2
 \end{equation}
\begin{equation}
     A_{torus} = \pi R^2 + \pi r^2 + 2\pi Rr(2\pi -1)
 \end{equation}
 \begin{equation}
     A_{spheroid} = 4\pi \left(\frac{R^{3.2} + 2(Rr)^{1.6}}{3}\right)^\frac{1}{1.6}
 \end{equation}
Using all these parameters, we turn to calculate the change in surface energy during the formation of cup, torus or spheroid shapes from initial disk-shaped silica.

\begin{figure}[h!]
\includegraphics[width=0.8\textwidth]{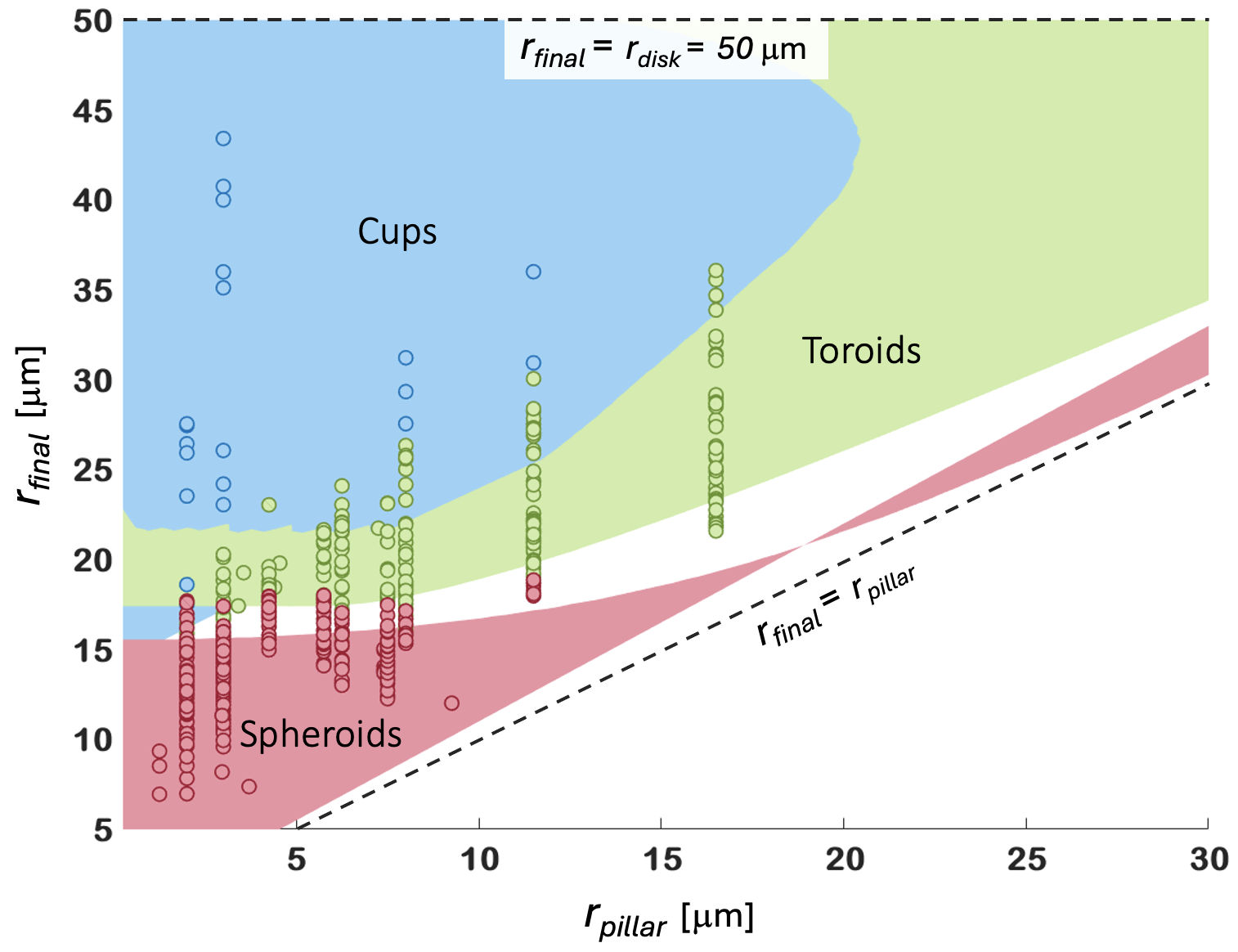}
\centering
\caption{Phase diagram of the resulting geometrical shapes after the reflow process, as a function of the initial pillar radius $r_{pillar}$ and final radius $r_{final}$, calculated according to minimal surface energy. The different geometrical structures are presented by colors: Cups (blue), toroids (green), ellipsoids (red). Shaded areas are the model predictions and the circles denote the experimentally measured samples.}
\label{fig:phase_dia}
\end{figure}

The expression for the change in the energy, namely the work $dW$, depends both on the change in the area and on the change in the curvature \cite{GURKOV199045}:
\begin{equation}
    dW = \gamma dA -\frac{2\gamma}{R_{curv}}AdR_{curv}
    \label{eq: surface energy minimization}
\end{equation}

where $\gamma$ is the silica's surface tension.\\
Assuming that for every set of $r_{pillar}$, $r_{final}$ the final shape will be the one maximizing the (negative) work, in figure \ref{fig:phase_dia} we present the phase diagram of the shapes that give the  minimal final surface energy (for the case of initial $r_{disk}=50\mu m$ and $w=2\mu m$).\\
Experimentally, for every $r_{pillar}$ we started with laser intensities that gave us the first noticeable folding, resulting in $r_{final}$ that are not far from the initial $r_{disk}$ of 50$\mu$m. As expected, using higher intensities yielded lower and lower $r_{final}$ (downwards in figure \ref{fig:phase_dia}), reaching values that are slightly above $r_{pillar}$. The possible final shapes are marked by the colored areas for the theory, and by the color of the circles for experimental results, with cups, toroids and spheroids marked by blue, green and red correspondingly. As evident, there is generally a very good agreement between the resulting shapes and the theoretical prediction based on minimal surface energy.

\begin{figure}[h!]
  \centering
  \includegraphics[width=0.9\linewidth]{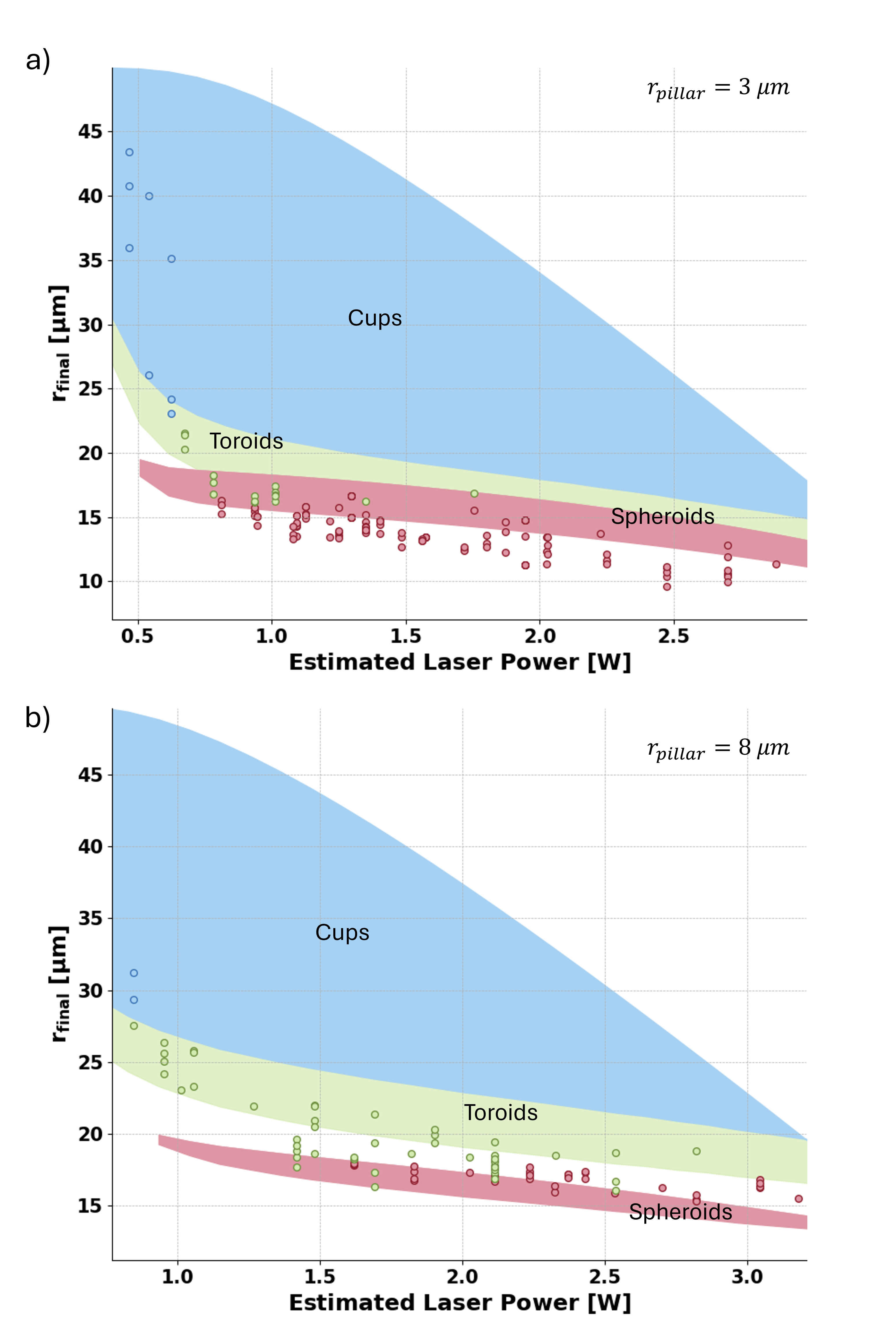}
  \caption{Iso-pillar radius slices of the phase-diagram representing \(r_{final}\) as function of the laser intensity. In red: Spheroids with ellipticity of 0-0.3; In green: Toroids; In Blue: Cups.}
\label{fig:rf_Ilaser}
\end{figure}

Looking at a representative iso-pillar radius slices of \(3\mu m\) and \(8\mu m\), figure \ref{fig:rf_Ilaser} depicts the additional aspect of the phase diagram where \(r_{final}\) is plotted as function of the laser intensity. According to the model, spheroids with ellipticity ranging between 0 to 0.3 can be produced in the red region and toroids and cups in the green and blue regions respectively.
The results align with the presented model, indicating that parameters such as the disk and pillar radii and the laser intensity can be set in advance to achieve the desired resonator geometry and dimensions. Therefore, the phase diagrams act as a valuable predictive tool or ‘map’ for achieving a specific resonator geometry.
However, it is important to note that it is always hard to attain the exact values of all the physical parameters involved. Beyond the empirical uncertainty in parameters such as absorption coefficients and thermal conduction of the specific materials at the temperature and wavelengths ranges involved, it is challenging to obtain exact values for other system parameters. In our case, the pillar radii, although measured with high accuracy using SEM, varied from one microresonator to the other (depending on the position within the chip). Additionally, the CO$_2$ laser power was measured after some optical elements (pinhole, mirrors), and also varied from shot to shot by at least 10$\%$-20$\%$. Accordingly, we treat the model and the results presented here only as semi-quantitative, and apply empiric adaptations to the measured parameters. Specifically, in the results presented in figure \ref{fig:rf_Ilaser}, we empirically fit the horizontal scale by assuming the laser power on the chip was 50$\%$ higher than measured by the detector after the pinhole and mirror. Additionally, the average temperature at the top of the silicon pillar was simulated by COMSOL simulations to be around 550K. The results of the model do not depend strongly on this value.

\newpage

\newpage
\section{Conclusion}
In this work, we present a detailed procedure for the fabrication of ultra-high Q chip-based silica WGM microresonators. Following this recipe, we were able to produce resonators with quality factors ranging from \(3\cdot10^7\) to \(1.7\cdot10^8\) at 780nm in air. In addition, we provide a predicting tool for the final geometry of the WGM microresonator given initial parameters of the silica disk and silicon pillar diameters and the intensity of the laser beam used for the reflow process. This model analyzes the thermodynamic behavior of the silica disk after being illuminated by a CO\(_2\) laser pulse beam. Taking into account all the heat dissipation channels of the system, with conduction being the main one, the model evaluates the temperature distribution along the radial direction of the disk. Consequently, the model determines the mobile portion of the silica disk to fold into one of three possible geometries: spheroids, toroids and cups. The shape into which the disk will fold is determined based on the minimization of the surface area. The ability to tailor the initial parameters in order to fabricate a resonator of desired geometry and dimensions should become a valuable tool in the challenging field of cavity-QED with WGM microresonators. 

\bibliography{manuscript}

\newpage
\section{Supplementary information}
Here, we show in more detail the mass conservation and the final radii expressions connecting the radius of mobility to the final radii of the various geometries of the WGM microresonatros.  

\subsection{Toroids}
The initial volume to be folded is:
\begin{equation}
    V_{initial} = w\cdot \pi (r_{disk}^2 - r_{mobility}^2) (1-p_{vap})
\end{equation}
where \(p_{vap}\) is the percentage of the vaporized volume.\\

\begin{figure}[h]
    \centering
    \includegraphics[width=1\textwidth]{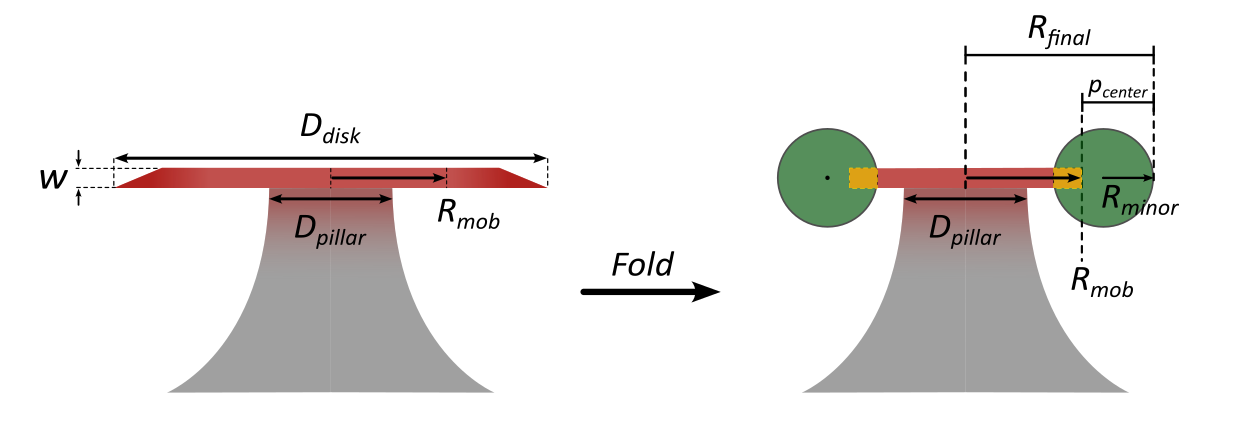}
    \caption{Schematic representation of the folding of a toroid microresonator. On the left: Initial geometry with initial parameters of width, disk and pillar diameters and radius of mobility (denoted as \(R_{mob}\)). On the right: Final toroid geometry. In yellow is the immobile portion of the disk overlapping with the circular cross-section (in the text denoted as \(V_{immobile}\)).}
    \label{fig:toroid}
\end{figure}

\noindent The final geometry of the toroid is:
\begin{equation}
    V_{toroid} = 2\pi^2(r_{mobility} + p_{center}\cdot 2r_{minor}-r_{minor})r_{minor}^2
\end{equation}
where \(p_{center}\) indicates the extent to which the circular cross section of the toroid protrudes out from the silica disk, relative to the mobility radius (see figure \ref{fig:toroid}).

Since the initial volume includes only the mobile portion to be folded, the immobile portion of the silica disk inside the torus shape needs to be subtracted as follows:
\begin{equation}
    V_{immobile} = w\cdot \pi (r_{mobility}^2 - (r_{mobility} - (1-p_{center})\cdot 2r_{minor})^2)
\end{equation}
\\
\noindent Thus, the mass conservation is as follows:
\begin{equation}
    (V_{toroid} - V_{immobile}) - V_{initial} = 0
\end{equation}
and the final radius is:
\begin{equation}
    r_{final, toroid} = r_{mobility} + p_{center}\cdot 2r_{minor} 
\end{equation}

\noindent Figure 7 in the main text depicts multiple possible toroidal geometries with \(p_{center}\) ranging from 30\(\%\) to 70\(\%\).

\subsection{Spheroids}
The initial volume is:
\begin{equation}
    V_{initial} = w \cdot \pi r_{disk}^2 (1-p_{vap})
\end{equation}

\noindent The volume of a spheroid with a certain amount of ellipticity is:
\begin{equation}
   V_{spheroid} = \frac{4}{3}\pi r_{long}^2 r_{short}
\end{equation}

\noindent where the relation between \(r_{long}\) and \(r_{short}\) is described by the ellipticity (flattening) factor:
\begin{equation}
    f_{ellipticity} = 1 - \frac{r_{short}}{r_{long}}
\end{equation}

\begin{figure}[h]
    \centering
    \includegraphics[width=1\textwidth]{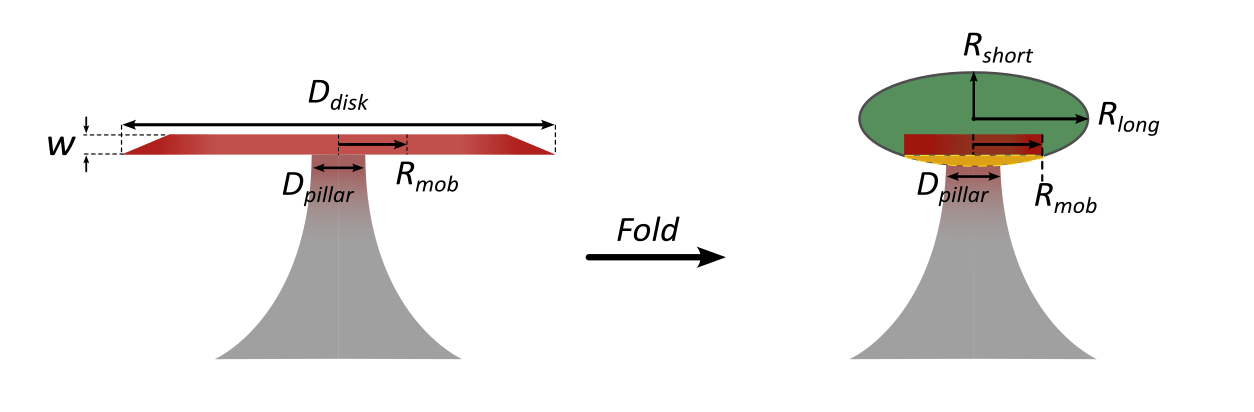}
    \caption{Schematic representation of the folding of a spheroid microresonator. On the left: Initial geometry with initial parameters of width, disk and pillar diameters and radius of mobility (denoted as \(R_{mob}\)). On the right: Final spheroid geometry with \(R_{long}\) as the final radius. In yellow is the spheroid cap being excluded from the final shape.}
    \label{fig:spheroid}
\end{figure}

\noindent In the calculation of the final spheroid geometry we need to exclude the volume segment overlapping with the pillar (see figure \ref{fig:spheroid}). This (relatively small) volume was neglected in the analysis presented in figure 6, yet for the detailed results presented in figure 7 we take it into account for higher accuracy. This segment is expressed by the volume of a spheroid cap:
\begin{equation}
   V_{cap} = \pi r_{long}^2 \left(\frac{2r_{short}}{3} - (r_{short} - L_{segment}) - \frac{(r_{short} - L_{segment})^3}{3r_{short}^2}\right)
\end{equation}

\noindent where,
\begin{equation}
    L_{segment} = r_{short}\left(1-\sqrt{1 - \left(\frac{r_{mobility}}{r_{long}}\right)^2}\right)
\end{equation}

\noindent Thus, the mass conservation for the spheroid final geometry is:
\begin{equation}
    (V_{spheroid} - V_{cap}) - V_{initial} = 0
\end{equation}

\noindent And the final radius is:
\begin{equation}
    r_{final,spheroid} = r_{long}
\end{equation}

\noindent Figure 7 in the main text depicts multiple possible spheroid geometries with ellipticity ranging from 0 (a sphere, i.e. \(r_{long} = r_{short}\)) to 0.3.

\subsection{Cups}
The analytical approach differs between figure 6 and figure 7 in the main text. Similar to the spheroids and toroids, The analysis of figure 6 incorporates the cup's final geometry, utilizing equations 3.13-3.15 for a comprehensive assessment of its surface energy. In contrast, figure 7 employs a simplified representation of the cups due to the following assumption: generally, the system will be driven to acquire the smallest surface area as possible from surface energy minimization reasons. Therefore, among the possible geometries, cups are less favorable and will only be formed under conditions that do not allow the formation of spheroids and toroids. Based on this assumption, the cups region in figure 7 is indicated wherever spheroids and toroids cannot form, taking into account the volume loss due to evaporation during the reflow process.

\newpage

\end{document}